\documentclass[prd,twocolumn,showpacs,preprintnumbers,amsmath,amssymb,superscriptaddress,amsmath]{revtex4-1}

\makeatletter
\renewcommand*\env@matrix[1][c]{\hskip -\arraycolsep
  \let\@ifnextchar\new@ifnextchar
  \array{*\c@MaxMatrixCols #1}}
\makeatother

\usepackage{amsmath,array}
\usepackage{graphicx,subfigure}  
\usepackage{dcolumn}                 
\usepackage{bm}                          
\usepackage{epstopdf}
\usepackage{soul}
\usepackage{footnote}
\begin{document}
 
\preprint{APS/123-QED}

\title{Wavelet Approach to Search for Sterile Neutrinos in Tritium $\beta$-Decay Spectra}


\author{S.\ Mertens}
\affiliation{Institute for Nuclear and Particle Astrophysics, Nuclear Science Division, Lawrence Berkeley National Laboratory, USA}
\affiliation{Institute for Nuclear Physics (IKP), Karlsruhe Institute of Technology, Germany}
\author{K.\ Dolde}
\affiliation{Institute for Nuclear Physics (IKP), Karlsruhe Institute of Technology, Germany}
\author{M.\ Korzeczek}
\affiliation{Institute for Nuclear Physics (IKP), Karlsruhe Institute of Technology, Germany}
\author{F.\ Glueck}
\affiliation{Institute for Nuclear Physics (IKP), Karlsruhe Institute of Technology, Germany}
\affiliation{Wigner Research Institute for Physics, P.\ O.\ B. 49, H-1525 Budapest, Hungary}
\author{S.\ Groh}
\affiliation{Institute for Nuclear Physics (IKP), Karlsruhe Institute of Technology, Germany}
\author{R.\ D.\ Martin}
\thanks{currently at \textit{Department of Physics, University of South Dakota, USA}}\affiliation{Institute for Nuclear and Particle Astrophysics, Nuclear Science Division, Lawrence Berkeley National Laboratory, USA}
\author{A.\ W.\ P.\ Poon}
\affiliation{Institute for Nuclear and Particle Astrophysics, Nuclear Science Division, Lawrence Berkeley National Laboratory, USA}
\author{M.\ Steidl}
\affiliation{Institute for Nuclear Physics (IKP), Karlsruhe Institute of Technology, Germany}

\date{\today}
\begin{abstract}
Sterile neutrinos in the mass range of a few keV are candidates for both cold and warm dark matter. An ad-mixture of a heavy neutrino mass eigenstate to the electron neutrino  would result in a minuscule distortion - a 'kink' - in a $\beta$-decay spectrum. In this paper we show that a wavelet transform is a very powerful shape analysis method to detect this signature. For a tritium source strength, similar to what is expected from the KATRIN experiment, a statistical sensitivity to active-to-sterile neutrino mixing down to $\sin^2 \theta= 10^{-6}$ ($90\%$ CL) can be obtained after 3 years of measurement time. It is demonstrated that the wavelet approach is largely insensitive to systematic effects that result in smooth spectral modifications. To make full use of this analysis technique a high resolution measurement (FWHM of $\sim100$~eV) of the tritium $\beta$-decay spectrum is required.
\end{abstract}

\maketitle

\section{Introduction}
Our current knowledge of particle physics comprises three light neutrino mass eigenstates $\nu_1$, $\nu_2$, and $\nu_3$, which mix and form the three flavor eigenstates $\nu_e$, $\nu_{\mu}$, and $\nu_{\tau}$. A number of observations, however, suggest the existence of sterile neutrino states. These would not interact at all with other particles in the Standard Model (SM), but could mix with the active neutrinos and form additional neutrino mass eigenstates~\cite{snu-intro, snu-whitepaper}.

For example, a natural way to incorporate neutrino masses into the SM is by introducing a right-handed counterpart to the purely left-handed active neutrinos. Right-handed neutrinos cannot couple to the weak force carriers, and hence can only be observed via their oscillation to left-handed neutrino states. 

Without violating gauge invariance of the SM, a Majorana mass term for the right-handed neutrino can be introduced. Since no electroweak symmetry breaking is required for this mass term (i.\ e.\ no Higgs mechanism), it can have an arbitrary scale. The puzzling fact that neutrinos are six orders of magnitude lighter than all other fermions can be explained by choosing a rather high scale for this mass term, naturally creating the known light neutrino mass eigenstates and additional very heavy neutrino mass eigenstates~\cite{neutrino-masses}.

Another observations motivating the existence of additional neutrino states are recent anomalies in reactor and short-baseline experiments. These might be resolved by the existence of additional light sterile neutrinos mass eigenstates with a mass of $\sim 1$~eV~\cite{lightsterile, Reactor, Gallium, MiniBooNE}. (The speed of light c, and the Planck constant $\hbar$ are set to unity in this paper.)

In this paper, we focus on sterile neutrinos in the mass range of a few keV, which are motivated by cosmological observations.
These neutrinos would be one of the most promising candidates for both warm (WDM) and cold dark matter (CDM)~\cite{SterileDM1, SterileDM2, SterileDM3, SterileDM4, SterileDM5, SterileDM6, SterileDM7}. The existence of WDM would reconcile recent structure observations from sub-galactic to larger scales. For example, the tension between the predicted and measured number of dwarf satellite  galaxies in the $\Lambda$CDM concordance model of cosmology can be mitigated by adding a sizable contribution of WDM~\cite{WDM1,WDM2,WDM3,WDM4,WDM5,WDM6,WDM7,WDM8,WDM9,WDM10,WDM11,WDM12,WDM13,WDM14,WDM15,WDM16}. 

Current cosmological observations limit the mass $m_\mathrm{s}$ and the active-to-sterile neutrino mixing angle $\sin^2\theta$ of keV-scale sterile neutrinos to 1 keV $< m_{\mathrm{s}} <$ 50 keV and $\sin^2\theta < 10^{-7}$~\cite{SterileConstraint1, SterileConstraint2, SterileConstraint3, SterileConstraint4, SterileConstraint5, SterileConstraint6}. Interestingly, recent analyses of XMM Newton telescope X-ray data show a very weak unidentified emission line from stacked galaxy cluster which could be possibly explained by the decay of a sterile neutrino with $m_{\mathrm{s}}=7.1~\mathrm{keV}$ and active-to-sterile neutrino mixing of $\sin^2\theta=7\cdot 10^{-11}$. Similar results have been reported for X-ray spectra of the Andromeda galaxy and the Perseus cluster~\cite{Bulbul,M31,Abazajian2014}.    

A small admixture of a keV-scale sterile neutrino mass eigenstate to the electron neutrino $\nu_\mathrm{e}$ would manifest itself in high resolution $\beta$ spectroscopy: At a specific energy below the endpoint corresponding to the keV-scale sterile neutrino mass $m_\mathrm{s}$ the reaction kinematics provide enough energy for emission of a heavy sterile neutrino mass eigenstate along with the electron. The opening of this new reaction channel thus results in a distinctive kink-like signature in the energy spectrum~\cite{Sch80, San13trit, Rod14, Rod14b}.

This paper details an investigation of using wavelet transforms to detect the spectral distortion from active-to-sterile neutrino mixing in tritium $\beta$ decays. It is shown that a statistical sensitivity to mixing angles $\sin^2 \theta \geq 10^{-6}$ (90\% CL) can be achieved with the statistics equivalent to three full beam years expected from a KATRIN-like $\beta$-decay experiment~\cite{designreport, katrin}. Most importantly, it demonstrates that the detection of the kink signature using wavelet transform is almost independent of the exact shape of the spectrum. It will be shown that a differential measurement with high energy resolution is necessary to reach this sensitivity.

\section{KeV-Scale Sterile Neutrinos and Tritium $\beta$-decay}
This section describes the imprint of keV-scale sterile neutrinos on the tritium $\beta$-decay spectrum. Furthermore, we give a brief introduction to the KATRIN (KArlsruhe TRItium Neutrino) experiment~\cite{designreport, katrin} and its strategy to measure the tritium $\beta$-decay spectrum in a narrow region close to the endpoint. We outline necessary modifications to the experiment in order to probe the region in the $\beta$-decay spectrum where the signature of active-to-sterile neutrino mixing can be expected.

\subsection{Imprint of a keV-Scale Sterile Neutrino on the $\beta$-decay Spectrum}
In the super-allowed $\beta$ decay of tritium
\begin{equation}
^3 \text{H} \rightarrow ~ ^3\text{He}^+ + e^- + \bar{\nu}_{\mathrm{e}} \text{ ,}
\label{3_TritiumDecay}
\end{equation}
an electron (anti-)neutrino eigenstate is created, which is a superposition of different mass eigenstates. Correspondingly, the $\beta$-decay spectrum is a superposition of spectra corresponding to the single mass eigenstates. Since the mass splittings between the known light mass eigenstates $m_{\mathrm{i}}$ are too small to be resolved experimentally, an 'effective electron neutrino' mass 
\begin{equation}
m^2_{\mathrm{light}}=\sum_i|U_{\mathrm{ei}}|^2 m_{\mathrm{i}}^2
\end{equation}
is measured, where $U_{\mathrm{ei}}$ represent elements of the PMNS matrix~\cite{PMNS}.

In the case where the electron neutrino contains an admixture of a neutrino mass eigenstate with a mass $m_{\mathrm{s}}$ in the keV range, the different mass eigenstates will no longer form one effective neutrino mass term. In this case, due to the large mass splitting, the superposition of the $\beta$-decay spectra corresponding to the light effective mass term $m_{\mathrm{light}}$ and the heavy mass eigenstate $m_{\mathrm{s}}$, can be detectable. The differential $\beta$-decay spectrum can be written as

\begin{equation}
\begin{split}
\left(\frac{d\Gamma}{dE_e}\right)=&\cos^2 \theta \left(\frac{d\Gamma}{dE_e}\right)_{m_{\mathrm{light}}}\Theta(E_0-E_e-m_{\mathrm{light}}) \\
 &+ \sin^2 \theta \left(\frac{d\Gamma}{dE_e}\right)_{m_{\mathrm{s}}}\Theta(E_0-E_e-m_{\mathrm{s}}) \text{,}
\end{split}
\label{sincos}
\end{equation}
where $E_e$ is the kinetic energy of the electron, $\sin^2 \theta$ denotes the mixing of the heavy mass state to the electron neutrino flavor state, and $E_0=18.6~\mathrm{keV}$ is the endpoint energy.

\begin{figure}[]
\centering
\includegraphics[width=0.45\textwidth]{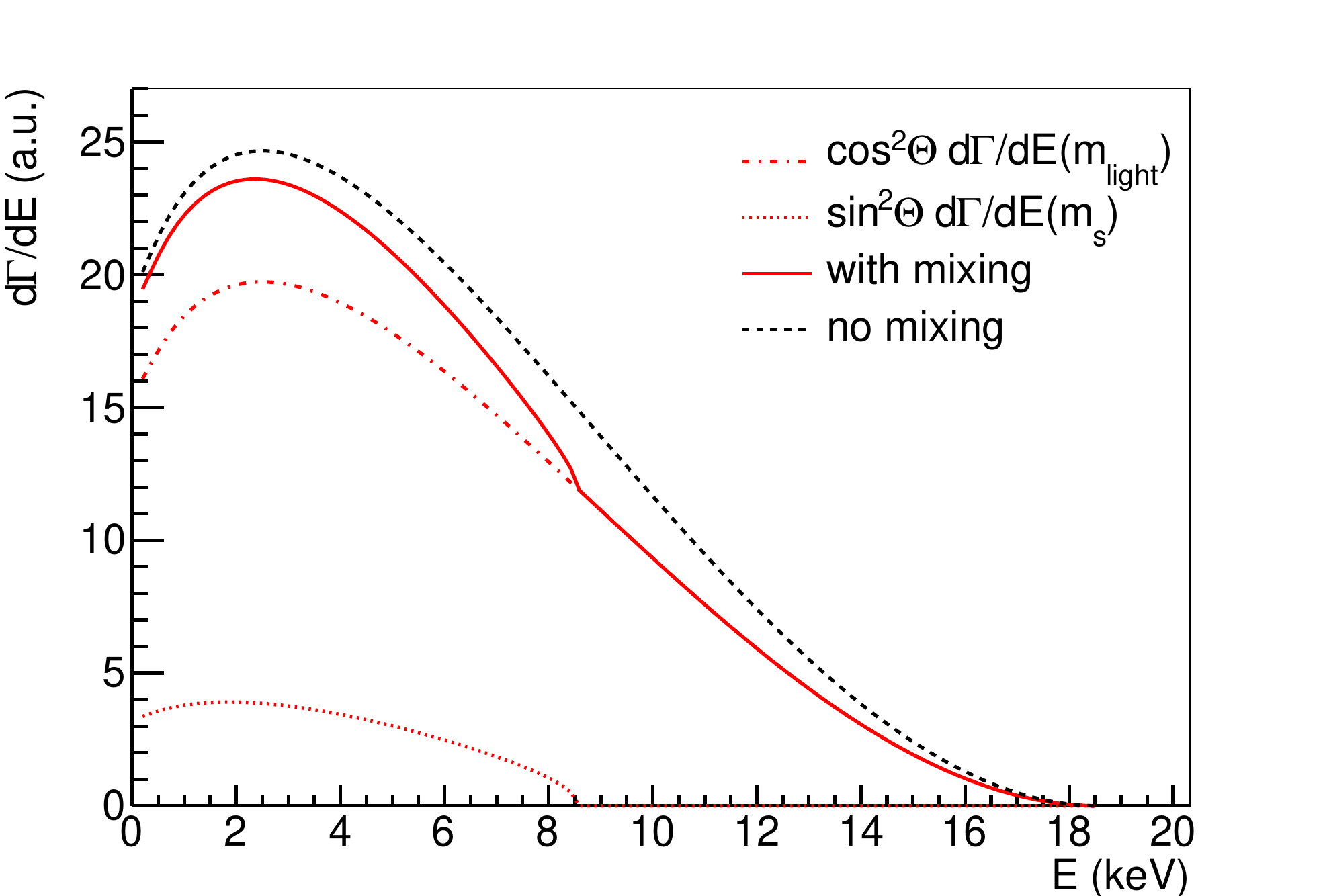}
\caption{Differential tritium $\beta$-decay spectrum without active-to-sterile neutrino mixing (black dashed line) and with mixing of a sterile neutrino with $m_{\mathrm{s}}=10$~keV and $\sin^2 \theta=0.2$ (red solid line). The red dashed lines depict the two components of the spectrum with active-to-sterile neutrino mixing.}
\label{fig:spectrum}
\end{figure}

Figure~\ref{fig:spectrum} shows the effect of active-to-sterile neutrino mixing with $\sin^2 \theta=0.2$ for a heavy sterile neutrino mass $m_{\mathrm{s}}=10$~keV in the tritium $\beta$-decay spectrum. The kink-like signature in the spectrum results from the superposition of individual differential $\beta$-decay spectra corresponding to light and heavy neutrino emission.

\subsection{Measurement of the Tritium $\beta$-decay Spectrum }
\label{ssc:measurement}
The KATRIN experiment is designed to measure the effective electron neutrino mass $m_{\mathrm{light}}$ with a sensitivity of 200~meV ($90\%$ CL)~\cite{designreport, katrin}. This will be achieved by an integral measurement of the tritium $\beta$ spectrum in the close vicinity to the endpoint where the influence of $m_{\mathrm{light}}$ is maximal. 

The $\beta$ source of KATRIN is an ultra-luminous windowless gaseous molecular tritium source (WGTS), which will deliver $\sim 10^{11}$ $\beta$ decays per second. A transport system guides the electrons from the WGTS to the spectrometer, while a system of differential pumping and cryotrapping prohibits tritium molecules from reaching the spectrometer. The spectrometer operates as a so-called MAC-E filter (\underline{M}agnetic \underline{A}diabatic \underline{C}ollimation combined with an \underline{E}lectrostatic Filter), providing an energy resolution of $<1$~eV. The basic idea is to set the spectrometer to a certain retarding potential, transmitting only those electrons with sufficient kinetic energy to overcome the potential barrier. By measuring the electron rate with a focal plane detector at the exit of the spectrometer for different retarding potentials, the integral $\beta$-spectrum is recorded.

The main advantage of KATRIN in view of a keV-scale sterile neutrino search is the tritium source strength~\cite{Bab12}, providing unprecedented statistics to allow probing small mixing angles. However, the existing segmented focal plane detector~\cite{focalplane}, as implemented for the low rate light neutrino mass measurement close to the endpoint, is not designed to handle the enormous rates of $\sim10^{11}$ counts per second.

A possible realization of a new detector system could be a highly modular detector array, comprising up to $\sim10^6$ pixels. A promising technology for high rate multi-pixel detectors could be based on Silicon Drift Detectors~\cite{Det1,Det2}.

This paper is not intended to outline a detailed technical realization of a new detector system. For our analyses we assume a generic detector system, capable of coping with the expected count rates, and we distinguish two generic measurement modes:

\textbf{Differential measurement mode:} Unlike in the usual KATRIN operating mode, the retarding potential would be set to a fixed low value at all times, transmitting the entire part of the $\beta$-decay spectrum of interest. In this case, we assume that the detector itself provides sufficient energy resolution to measure the differential tritium $\beta$-decay spectrum. 

\textbf{Integral measurement mode:} As in the usual KATRIN operation mode, this measurement mode makes use of the unprecedented energy resolution in the eV range of the main spectrometer. In this case, the spectrum would be scanned by cyclic variation of the retarding potential of the main spectrometer. The detector system would measure the counting rate of the transmitted electrons and would not require a high energy resolution.

In the following we investigate the performance of a wavelet analysis of the tritium $\beta$-decay spectrum in detecting the kink-like signature of a keV-scale sterile neutrino. We consider the two described measurement modes, different energy resolutions, and the effect of uncertainties on the spectral shape.

\section{Wavelet Transform}
There is a wealth of literature on using different transformation techniques in signal processing. One of the most well-known methods is the Fourier transform, which transforms a time-dependent signal into the frequency domain. The Fourier transform provides a constant resolution for all frequencies depending on the signal length. Although the frequency components in the signal are known after a Fourier transform, there is no information on the point in time where a certain frequency appears in the signal.

In contrast, a wavelet transform is a technique that offers both time and frequency resolution~\cite{Wavelet1,Wavelet2,Wavelet3,Wavelet4,Wavelet5,Wavelet6,Wavelet7}. A Wavelet transform performs a convolution of specific window functions, so-called wavelets, with the given signal. In contrast to the sinusoidal basis functions of the Fourier transform, wavelets offer the major advantage of being localized in time and frequency. Changing the wavelet's position in the transformation of the signal provides the time resolution, while modulation of the wavelet length leads to the frequency resolution. 

In this section a short overview of the continuous and the discrete wavelet transform, which is the basis for the analysis in this paper, is presented.

\subsection{Continuous Wavelet Transform}
The continuous wavelet transform (CWT) of a continuous function $f(t)$ is defined as
\begin{equation}
\mathrm{CWT}(s,\tau)=\int f(t)\Psi_{s,\tau}^*(t)\mathrm{d}t.
\label{eq_CWT}
\end{equation}
In this notation~\cite{Wavelet7}, $\Psi_{s,\tau}(t)$ is the set of wavelets with the \emph{scale} parameter $s$ , the \emph{translation} parameter $\tau$. 

The wavelets are calculated from a \emph{mother wavelet} $\Psi_0(t)$. The set of wavelets is given by
\begin{equation}
\Psi_{s,\tau}(t)=\frac{1}{\sqrt{s}}\Psi_0\left(\frac{t-\tau}{s}\right),
\label{eq_wavelet_cwt}
\end{equation}
where scale parameter $s$ describes the size and thereby the frequency band of the wavelet. Small scales refer to high frequencies and large scales refer to small frequencies. The translation parameter $\tau$ shifts the wavelet relatively to the signal allowing an examination at different points of time. Thus, the CWT of a signal provides a two-dimensional spectrum with time (translation) and frequency (scale) resolution.

\subsection{Discrete Wavelet Transform}
\label{ssc:DWT}
To compute the discrete wavelet transform (DWT), several successive high-pass and low-pass filters are applied to a discrete time-domain signal. In the first step, the high-pass filter determines the amount (Power) of high frequencies (corresponding to ``scale 1'') at certain times in the signal. Scale 1 covers the frequency range from half the maximal frequency $\frac{f_{\mathrm{max}}}{2}$ to the maximal frequency $f_{\mathrm{max}}$. The low-pass filter eliminates these high frequencies and returns a smoothed signal. 

\begin{figure}
\centering
\includegraphics[width=0.45\textwidth]{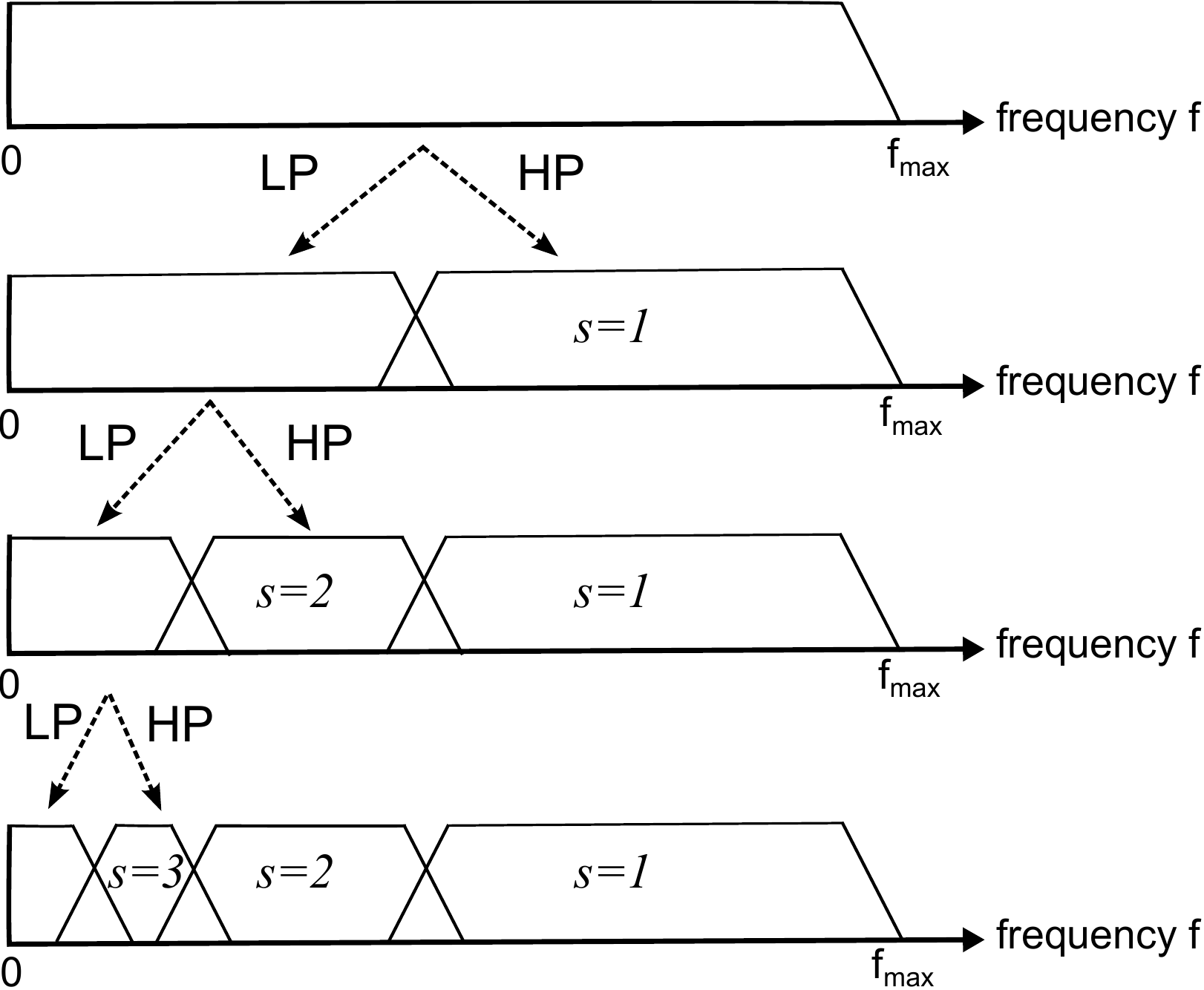} 
\caption{Schematic of sub-band coding, by applying successive high-passes (HP) and low-passes (LP) on the data. The frequency resolution increases with increasing scales.}
\label{fig:subbandcoding}
\end{figure}

In the second step, the wavelet transform acts on the smoothed signal: The high-pass filter determines the amount of frequencies, corresponding to ``scale 2'', in the smoothed signal and the low-pass filter subtracts these and returns a further smoothed signal. Scale 2 covers a smaller frequency range than scale 1, namely $\frac{f_{\mathrm{max}}}{4}$ to $\frac{f_{\mathrm{max}}}{2}$. This \emph{pyramidic algorithm}, visualized in figure~\ref{fig:subbandcoding}, is repeated successively until the signal is reduced to only two values. 

By applying this strategy of sub-band coding, the frequency resolution increases by a factor of two in each step while the time resolution decreases by the same factor. Hence, DWT has a good frequency resolution for low frequencies and a good time resolution for high frequencies. 

\begin{figure}
\centering
\includegraphics[width=0.45\textwidth]{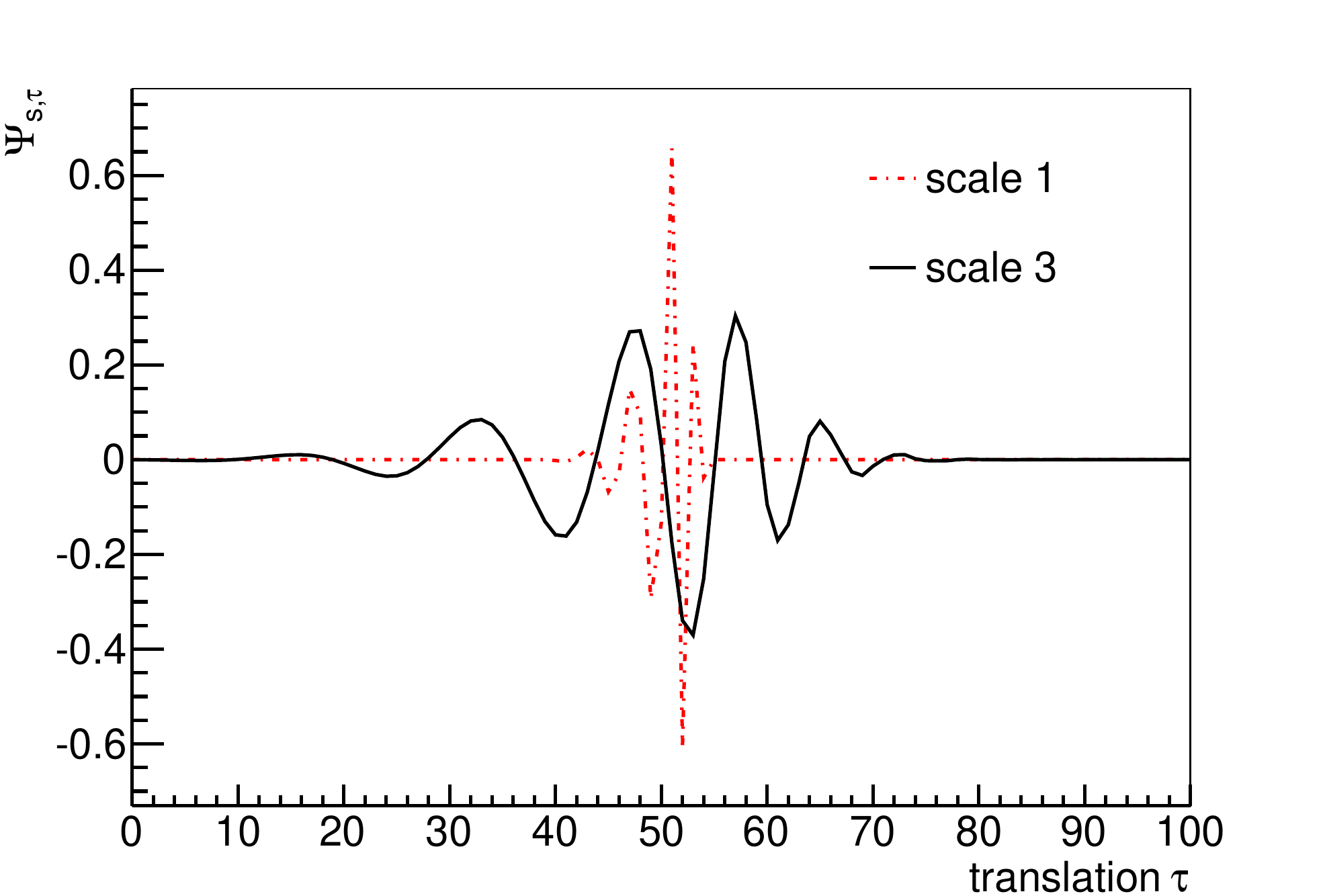}
\caption{Comparison of two different scales (1,3) of the mother wavelet \emph{Daubechie 18}. The wavelet of scale 1 covers a rather large frequency range (see figure~\ref{fig:subbandcoding}), is however very localized in $\tau$. Wavelets of larger scale cover smaller frequency ranges, but are less localized in $\tau$.}
\label{fig:wavelet}
\end{figure}

In our analysis, we implemented the DWT as a matrix transformation~\cite{Wavelet3} performing the successive high- and low-pass filtering on the data vector. In our case, the data vector describes the tritium $\beta$-decay spectrum from $0-18.6$~keV with 512 points. The coefficients of this matrix are specific to a certain mother function \emph{Daubechie n}~\cite{Wavelet1,Wavelet2}. Our investigations showed that n=18 (as displayed in figure~\ref{fig:wavelet}) is most efficient for our purposes.

\section{Wavelet Analysis of the Tritium $\beta$-decay Spectrum}
In the first part of this section we demonstrate the properties of wavelet transform in analyzing tritium $\beta$-decay spectra to search for the signature of a keV-scale sterile neutrino. In the second part we introduce the statistical procedure to quantify the sensitivity of our analysis technique. 

\subsection{Technical Realization of the DWT}
The tritium $\beta$-decay spectrum is described by a data vector $x_i$ of length $N=512$, where $x_i$ is the decay rate in the energy bin $i$. In the first step of the DWT  this data vector is transformed by the $N\times N$ transformation matrix $T^{s=1}$ to obtain the power values $\gamma^{s=1}$ and smoothed data values $x^{s=1}$. $\gamma^{s}$ and $x^{s}$ of all scales s ($1\leq s \leq \frac{\ln N}{\ln 2}$) are obtained by recursively applying the transformation matrix $T^{s}$ to the smoothed data vector $x^{s-1}$ 
\begin{eqnarray}
\gamma^{s}_k &=& \sum_{i=1}^{N^{s}} T_{2k,i}^{s} x_i^{s-1},\\
x_k^{s} &=& \sum_{i=1}^{N^{s}} T_{2k-1,i}^{s} x_i^{s-1},
\end{eqnarray}
where $1 \leq k \leq \frac{N^{s}}{2}$.
The entries in odd rows of this matrix are scaling coefficients, acting as a low-pass filter, the even entries are so-called wavelet coefficients and operate as the high-pass filter on the data vector. We choose our parametrization according to~\cite{Wavelet3}. 

\begin{figure}[]
\centering
\includegraphics[width=0.45\textwidth]{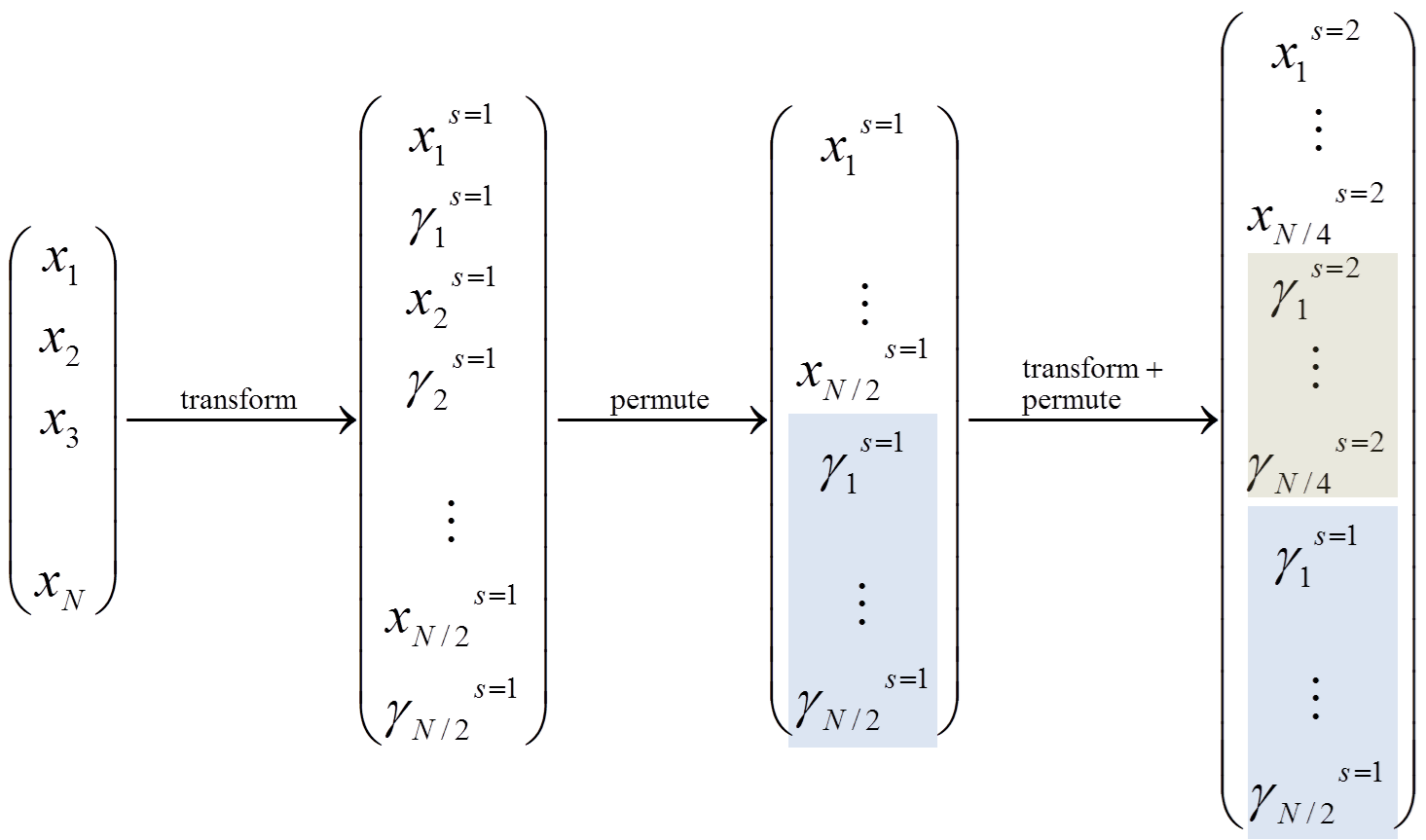}
\caption{Procedure of the DWT applied to a data vector $x_i$. $x_i^s$ denote the smoothed data, and $\gamma_i^{s}$ depict the power in a certain frequency band corresponding to scale $s$, as described in the main text. The blue and brown box indicate the $s=1$ and $s=2$ power values, respectively.}
\label{fig:transform}
\end{figure}

After the initial step, the resulting vector has $\frac{N}{2}$ smoothed data values $x^{s=1}_k$, and $\frac{N}{2}$ power values $\gamma^{s=1}_k$ of the smallest scale, i.\ e.\ $s=1$. In the next step, a reduced ($\frac{N}{2}\times \frac{N}{2}$) transformation matrix is applied to the smoothed data vector, providing $\frac{N}{4}$ twice-smoothed signal components $x^{s=2}_k$, and $\frac{N}{4}$ power values $\gamma^{s=2}_k$. This is repeated successively, until the length of the data vector is reduced to 2. Figure~\ref{fig:transform} displays the procedure.

\begin{figure}[]
\centering
\includegraphics[width=0.45\textwidth]{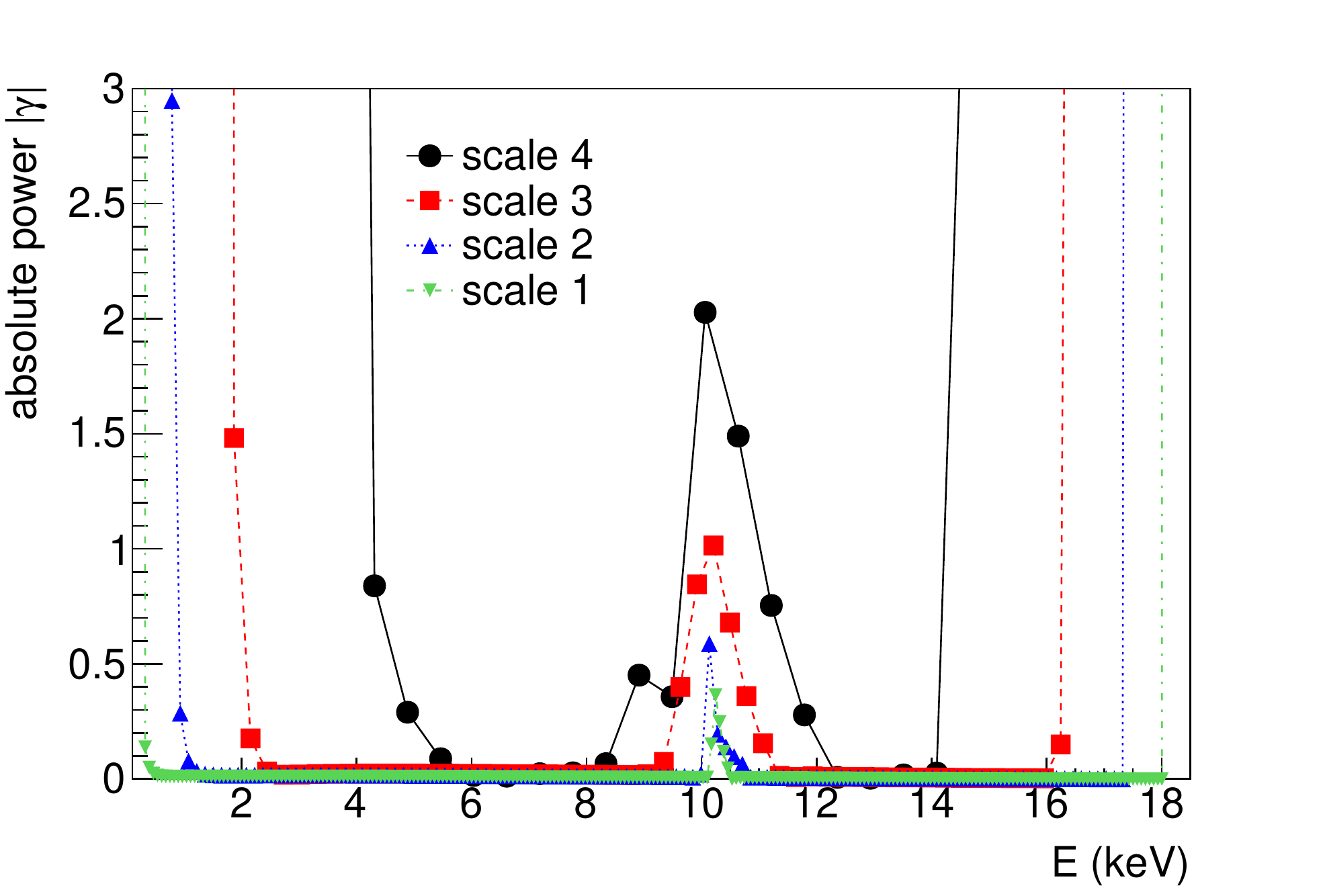}
\caption{Power spectrum of a $\beta$-decay spectrum for scales 1 - 4 with $m_{\mathrm{s}} = 8$~keV and $\sin^2\theta = 10^{-5}$. The wavelengths $L_s$ of the wavelets of different scales in units of energy bins are given by $L_s = 2^s$, corresponding to $\sim140$~eV (scale 2), $\sim280$~eV (scale 3), and $\sim562$~eV (scale 4). At the low- and high-energy part of the spectrum, boundary effects occur due to the non-periodicity of the tritium $\beta$-decay spectrum.}
\label{fig:powerspectrum}
\end{figure}

The power values $\gamma^{s}_k$ contain the information about the appearance of certain frequencies (corresponding to scale $s$) in the signal at position $k$ (i.\ e. at a certain $\beta$ energy in our case). The kink-like signature of a keV-scale sterile neutrino in the $\beta$-decay spectrum, see figure~\ref{fig:spectrum}, manifests itself in the power spectrum as a large excess of $\gamma^{s}_k$ at the kink position for multiple scales. Figure~\ref{fig:powerspectrum} displays $\gamma^{s}_k$ for four different scales, corresponding to the DWT of a tritium $\beta$-decay spectrum including the signature of a keV-scale sterile neutrino with a mass of $m_{\mathrm{s}}=8$~keV and a mixing angle of $\sin^2\theta = 10^{-5}$.

At the low- and high-energy ends of the power spectrum, boundary effects of the applied wavelet method constrain the sensitivity for keV-scale sterile neutrinos. This behavior arises from the fact that the wavelet transform is not optimized for non-periodic signals such as the tritium $\beta$-decay spectrum. In other words, a periodic continuation of the tritium $\beta$-decay spectrum, leads to a large step at the boundary. 

A possible way to reduce this effect is by modifying the spectral shape to allow for a smooth connection of the end to the beginning of the spectrum or to use a modified wavelet approach with a purely band-diagonal DWT matrix as described in~\cite{modifymatrix}. In this work we focus on the fully model-independent approach, based on the standard DWT of the unmodified spectrum.  

\subsection{Statistical Analysis Method}
To separate spectra with a kink signature from spectra without, we define the parameter $\Omega^s(m_{\mathrm{s}})$ as the sum of the absolute power values $|\gamma^{s}|$ in a window centered at the position $E_{\mathrm{kink}}$ corresponding to the keV-scale sterile neutrino mass we are probing:     
\begin{equation}
\mathrm{\Omega}^s(m_{\mathrm{s}})=\sum_{k=E_{\mathrm{kink}}-\Delta_s}^{E_{\mathrm{kink}}+\Delta_s}|\gamma^{s}_k|,
\end{equation}
where $\Delta_s$ is chosen as 5 times the length of the wavelet of the considered scale $s$.  

The power $|\gamma|$ depends on the relative position of $E_{\mathrm{kink}}$ within the wavelet. To avoid an unphysical dependence on the phase shift of the wavelet, we average the result over all integer shifts $\varphi_s=0\ldots L_s$, where $L_s = 2^s$ is the length of a wavelet in units of energy bins. We use 
\begin{equation}
\langle\mathrm{\Omega}^s(m_{\mathrm{s}})\rangle=\frac{1}{L_s}\sum_{\varphi_s = 0}^{L_s}\mathrm{\Omega}^s(m_{\mathrm{s}})
\label{eq:DV}
\end{equation}
as signal-sensing parameter in our analysis.

\begin{figure}[]
\centering
\includegraphics[width=0.45\textwidth]{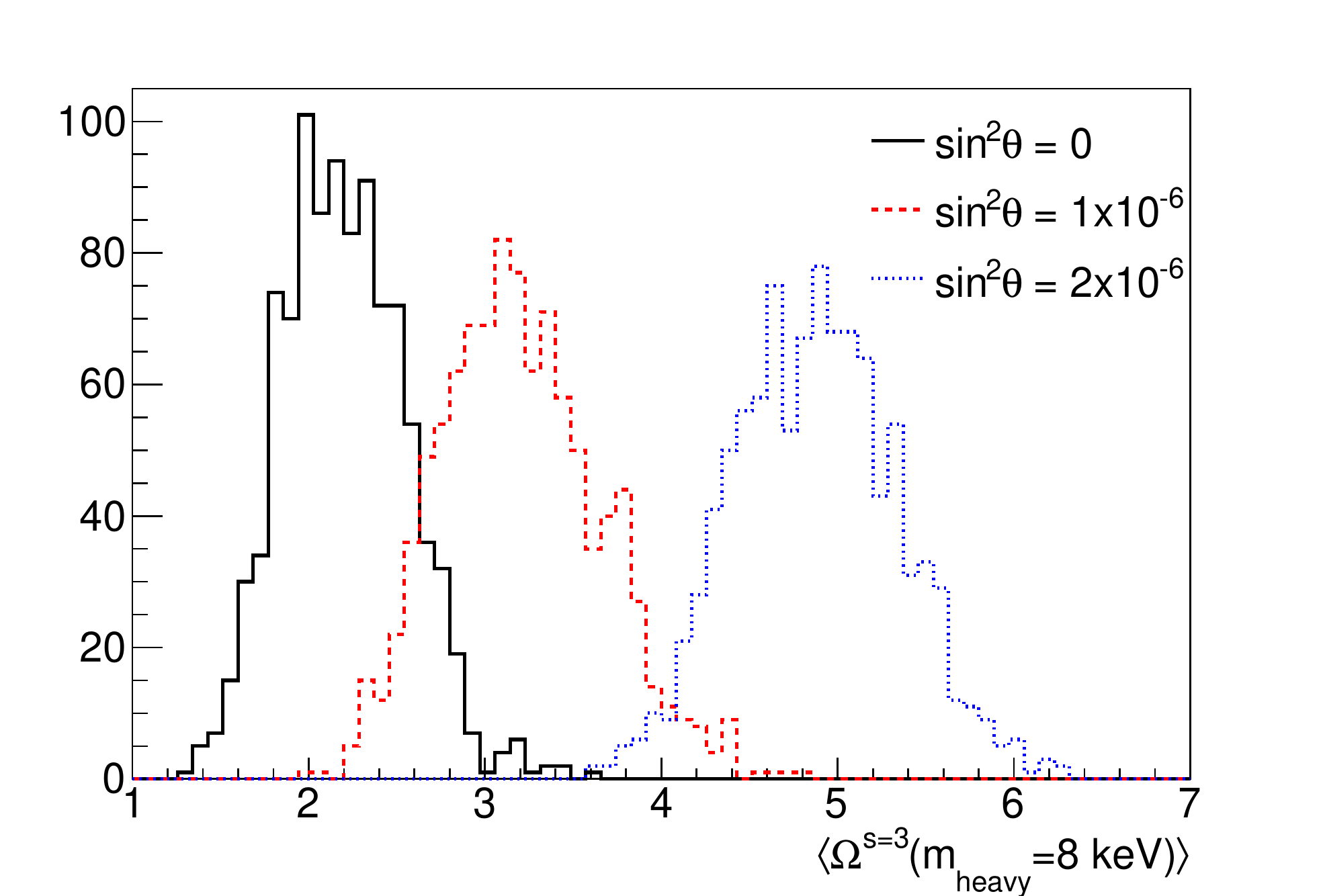}
\caption{Distribution of $\langle\mathrm{\Omega}^{s=3}(m_{\mathrm{s}}=8~\mathrm{kV})\rangle$ (see equation~\ref{eq:DV}) for $\sin^2\theta = 0$ (solid black line), $\sin^2\theta = 10^{-6}$ (red dashed line) and $2\cdot10^{-6}$ (blue dotted line). Each distribution is obtained from 1000 Monte Carlo simulations of the $\beta$-decay spectra assuming a total statistics of $~10^{18}$ electrons. This is achieved after 3-years differential measurement with the expected KATRIN source strength, taking into account angular acceptance and a generic 90\% detection efficiency.} 
\label{fig:MC}
\end{figure}

Figure \ref{fig:MC} shows the distributions of $\langle\mathrm{\Omega}^{s}(m_{\mathrm{s}})\rangle$ obtained from MC simulations for scale 3 and a keV-scale sterile neutrino of $m_{\mathrm{s}}=8$~keV.  We compare the distributions for the no-mixing case ($\sin^2\theta = 0$), to small mixing amplitudes of $\sin^2\theta = 10^{-6}$ and $2\cdot10^{-6}$, respectively. Assuming a source strength equivalent to what can be expected from KATRIN and a measurement time of 3 years a spectrum including active-to-sterile neutrino mixing with a mixing amplitude of $\sin^2\theta \approx 10^{-6}$ can clearly be distinguished from an undistorted spectrum.  

\begin{figure}[]
\centering
\includegraphics[width=0.45\textwidth]{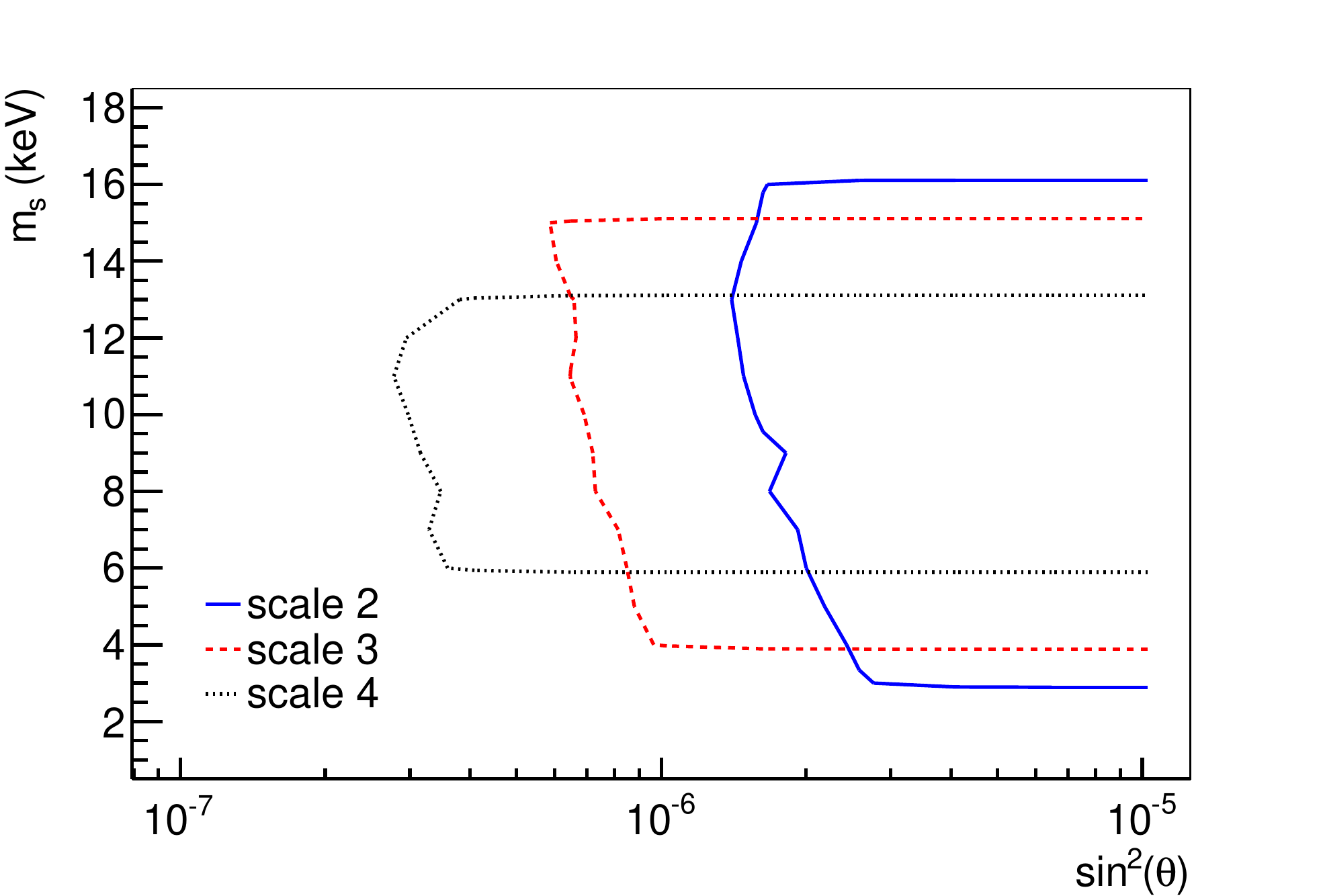}
\caption{90\% exclusion limits for total statistics of $~10^{18}$ electrons. Scale 4 reaches the strongest exclusion limit, however, boundary effects increase with increasing scale and limit the sensitivity in the low and high mass region.}
\label{fig:contour}
\end{figure}

Based on the basic principle in~\cite{Feldman}, we define the exclusion limit at $90\%$ confidence level, as the mixing angle $\sin^2\theta_{\mathrm{limit}}$, for which 90\% of the values of $\langle\mathrm{\Omega}^s(m_{\mathrm{s}})\rangle$ are larger than the expectation value of the $\langle\mathrm{\Omega}^s(m_{\mathrm{s}})\rangle$ distribution without active-to-sterile neutrino mixing ($\sin^2\theta = 0$). By determining $\sin^2\theta_{\mathrm{limit}}$ for each keV-scale sterile neutrino mass we can draw an exclusion contour in the ($m_{\mathrm{s}}$, $\sin^2\theta$) plane as shown in figure~\ref{fig:contour}.

Prior to applying the wavelet transformation to an actual measured spectrum, a 90\% C.L. cut value $\langle\mathrm{\Omega}^{s,\text{limit}}(m_{\mathrm{s}})\rangle$ will be determined based on simulation for each mass and scale. If the measured values $\langle\mathrm{\Omega}^{s,\text{meas}}(m_{\mathrm{s}})\rangle$ do not exceed these limits, a 90\% exclusion limit can be set. In case a positive signal is found, the Look-Elsewhere effect~\cite{LookElsewhere} needs to be taken into account, which assures a proper calculation of the significance level to find a kink-like signature when searching at serveral positions in the tritium spectrum.

\section{Sensitivity of the Wavelet Approach}
This section discusses the sensitivity of the wavelet technique including statistical and systematical effects. We compare the purely statistical sensitivity for the differential and integral measurement mode, as described in section~\ref{ssc:measurement}. Furthermore, we investigate systematic effects in a generic way. First, we demonstrate that the wavelet approach is largely insensitive to the precise shape of the tritium $\beta$-decay spectrum. Second, we show that an energy resolution of the order of about 100~eV is necessary in order to utilize the full potential of the wavelet approach.

\subsection{Statistical Sensitivity for Different Measurement Techniques}
\label{sec:satsensi}
\begin{figure}[]
\centering
\includegraphics[width=0.45\textwidth]{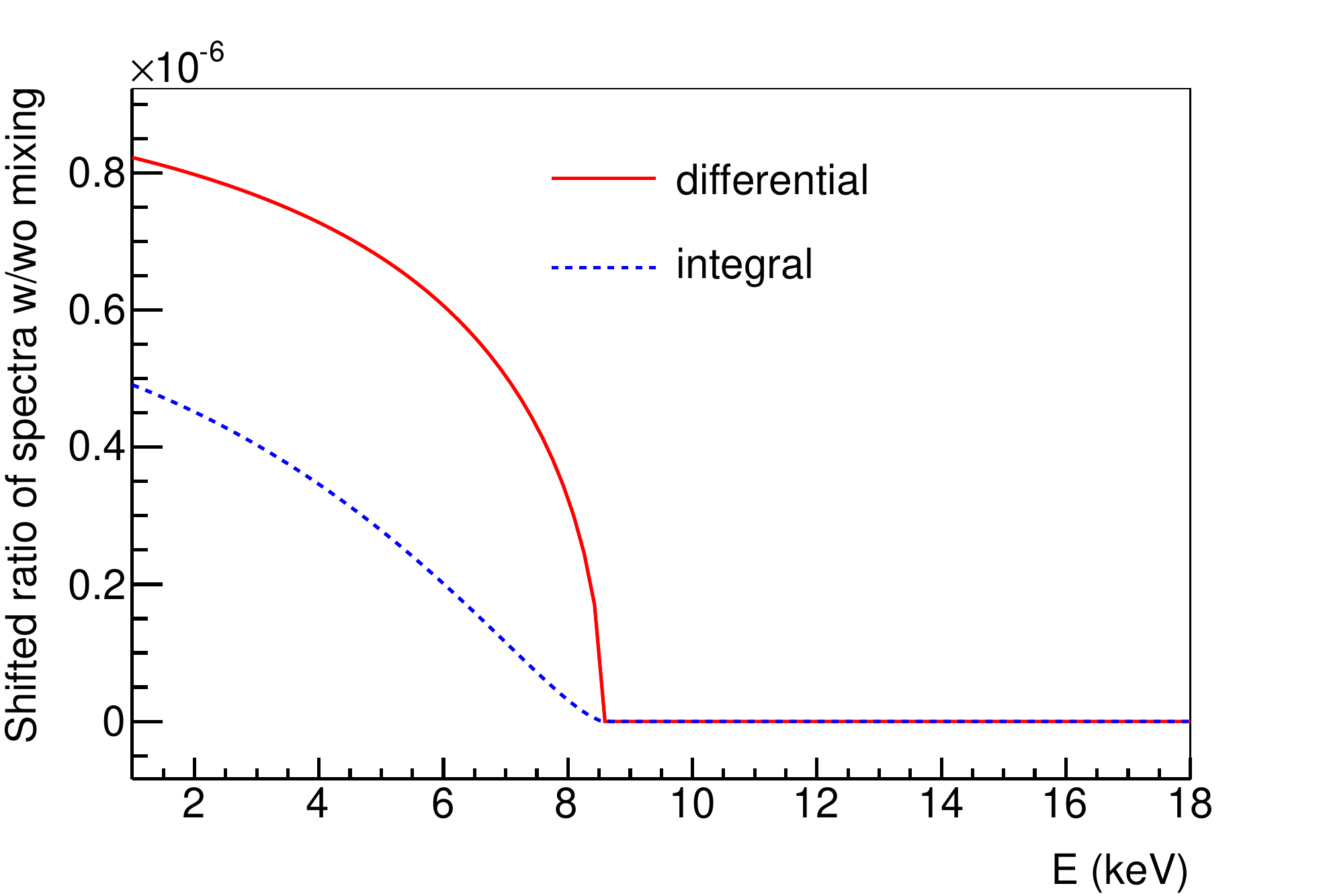}
\caption{Shifted ratio of the distorted and the undistorted tritium $\beta$-decay spectrum for both differential and integral measurements with $m_{\mathrm{s}}=10$ keV and a mixing angle of $\sin^2 \theta = 10^{-6}$. For easier comparison the curves have been shifted to 0 beyond the kink. The integral spectrum was calculated by assuming equal measurement times at 512 equally distributed retarding potentials of the main spectrometer.}
\label{fig:integral}
\end{figure}

To visualize the impact of different measurement techniques, the signature of a keV-scale sterile neutrino in an integral and a differential spectrum is shown in figure \ref{fig:integral}. The relative magnitude of the keV-scale sterile neutrino signature in the differential $\beta$-decay spectrum exceeds the one in the integral spectrum, resulting in a steeper slope next to the kink. Consequently, we expect a higher sensitivity in the case of a differential measurement (in addition to its inherent better statistics). 

\begin{figure}[]
\centering
\includegraphics[width=0.45\textwidth]{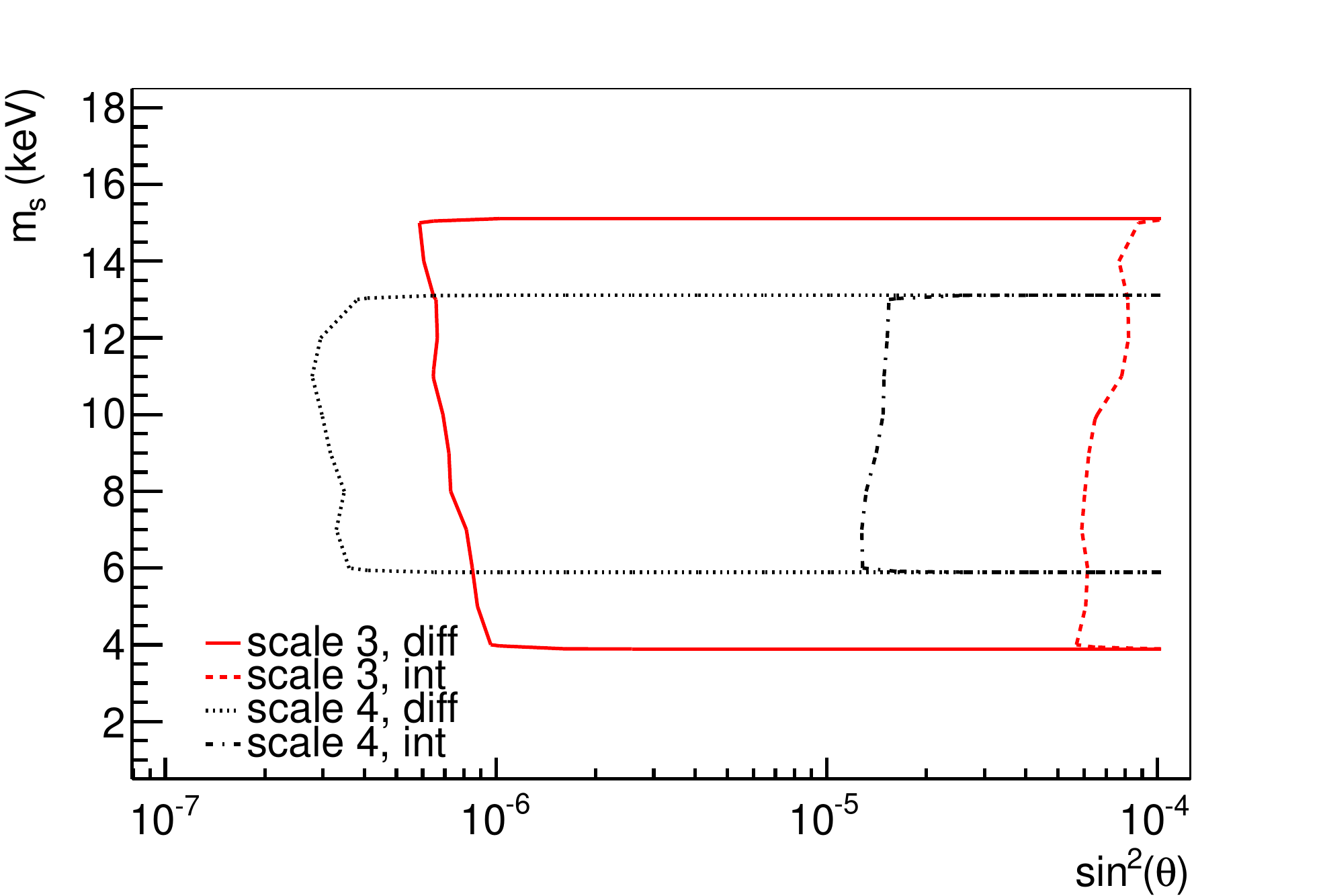}
\caption{Exclusion curves at $90\%$ CL for differential and integral spectra assuming a measurement time of 3 years with the expected KATRIN source strength. The analysis of a differential spectrum in scale 4 (black dotted lines) yields the best sensitivity in terms of $\sin^2\theta$. In scale 3 (red solid line) a larger mass range is covered but the sensitivity in terms of $\sin^2\theta$ is reduced. In both scale 3 (red dashed line) and 4 (back dotted dashed line), the sensitivity is drastically reduced, when applying the DWT to the integral spectrum as compared to the differential spectrum.}
\label{fig:exclusion_int}
\end{figure}

Figure \ref{fig:exclusion_int} compares the sensitivity of an integral and differential measurement mode for different scales. For scale 4 a sensitivity of up to $\sin^2\theta > 4\cdot10^{-7}$ can be reached. Boundary effects limit the mass range to $m_{\mathrm{s}} = 6 - 13$~keV. The same scale applied to an integral spectrum reaches only $\sin^2\theta > 1.6\cdot10^{-5}$. With scale 3 a larger mass range of $m_{\mathrm{s}} = 4 - 15$~keV can be covered. The sensitivity in this case reaches $\sin^2\theta > 1\cdot10^{-6}$ for the differential measurement mode and $\sin^2\theta > 6.3\cdot10^{-4}$ for the integral mode. We conclude that a differential measurement mode is strongly preferable to exploit the full potential of the wavelet technique. 

\subsection{Impact of Systematical Uncertainties}
In this section, the systematic effect of theoretical uncertainty on the spectral shape and a finite energy resolution are investigated. Both effects are implemented in a generic way, by introducing a polynomial shape factor and a Gaussian smearing of the energy, respectively.

\subsubsection{Spectral Shape}
In standard fitting procedures of a $\beta$-decay spectrum, a precise knowledge of the spectral shape is necessary~\cite{keV}. In this section we demonstrate that the wavelet approach is largely independent of the spectral shape and therefore insensitive to spectral uncertainties, provided that corrections to the tritium $\beta$-decay spectrum are smooth functions of energy. 

Theoretical corrections arise on the particle, nuclear, atomic and molecular level. An ensemble of corrections has to be applied to the Fermi theory of $\beta$-decay. The most dominant comprise electron screening~\cite{Wilkinson}, radiative corrections~\cite{Radiative}, and electron exchange~\cite{Exchange}. The size of these corrections reaches 0.01\% at the $\beta$-spectrum thresholds. All these corrections change the shape of the tritium $\beta$-decay spectrum in a smooth way. 

The final state distribution (FSD) is anticipated to be the largest and, at the same time, least known effect. Only 43\% of all $^{3}$H decay to the electronic ground state. Furthermore, as KATRIN is providing a molecular tritium source, this ground state is broadened by rotational and vibrational eigenstates of the daughter molecule. As the FSD is a relevant systematic effect for the light neutrino mass measurement, it has been computed precisely for a region close to the endpoint. However, no calculation is currently available for $\beta$-energies further away from the endpoint, as would be needed for a keV-scale sterile neutrino search. Provided that the excitations probabilities change smoothly as a function of $\beta$ energy, the FSD would lead to large corrections close to the endpoint and a smooth correction further than 300~eV (corresponding to the maximal excitation energy) from the endpoint. 

In addition, from the experimental point of view, source and detector related systematic effects will alter the spectral shape. These mainly entail energy losses due to inelastic scattering of the electrons while leaving the source, an energy-dependent detection efficiency, due to energy loss in the detector dead-layer and backscattering of electrons from the detector surface.

\begin{figure}
\centering
\includegraphics[width=0.45\textwidth]{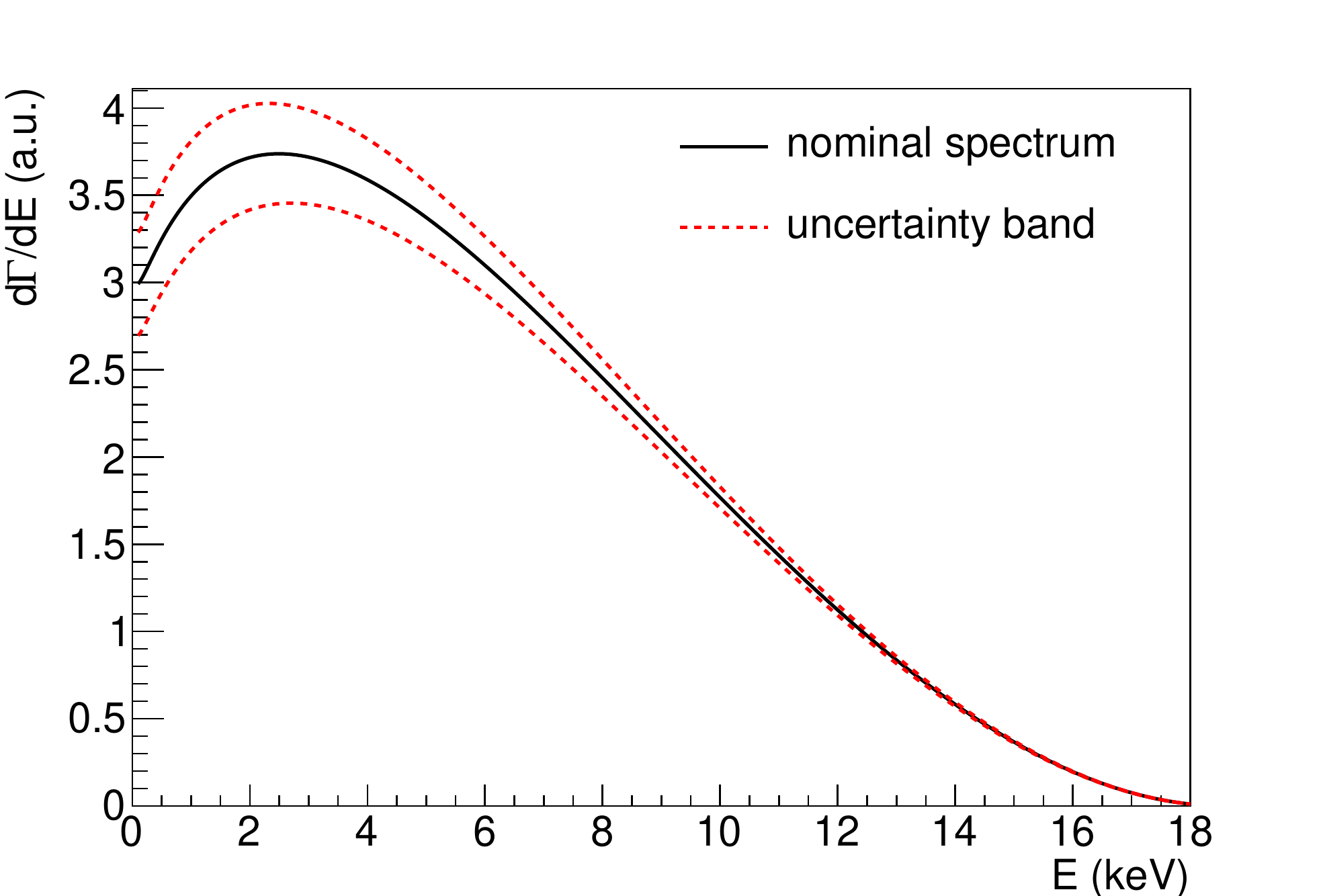} 
\caption{This figure illustrates the assumed 1$\sigma$ shape uncertainty of the tritium uncertainty. The red dashed lines depict the spectrum multiplied with a polynomial function (see equation~\ref{eq:polynom} where the linear, quadratic and cubic (a, b, c) coefficients are set to $\sigma = \pm10$\%, respectively. In the simulation the tritium $\beta$-decay spectrum is multiplied with an arbitrary polynomial, where a, b, and c are drawn from a Gaussian distribution with a width of $\sigma$.}
\label{fig:specband}
\end{figure}

Here, we test the wavelet technique with a heuristic approach: We incorporate smooth spectral shape corrections by multiplying the spectrum with a polynomial function of energy.
\begin{equation}
p(x)=  1+a\cdot x  + b\cdot x^2 + c \cdot x^3\text{,}~~ x = \frac{E_e-E_\text{0}}{E_\text{0}},   
\label{eq:polynom}
\end{equation}
where $E$ is the kinetic energy of the electron and $E_\text{0}$ is the spectrum endpoint. As the effort of understanding the spectral shape has been focused on a region close to the endpoint, we assume the uncertainties to increase towards lower $\beta$-electron energies. To simulate the effect of a spectral shape uncertainty on the sensitivity, we allow the spectral shape to vary within a band of the width $\sigma$, depicted in figure~\ref{fig:specband}. This is realized by drawing the parameters a, b, and c from a Gaussian distribution with a mean of zero and standard deviation $\sigma$, for each Monte Carlo simulations of the tritium $\beta$-decay spectrum. 

\begin{figure}
\centering
\includegraphics[width=0.45\textwidth]{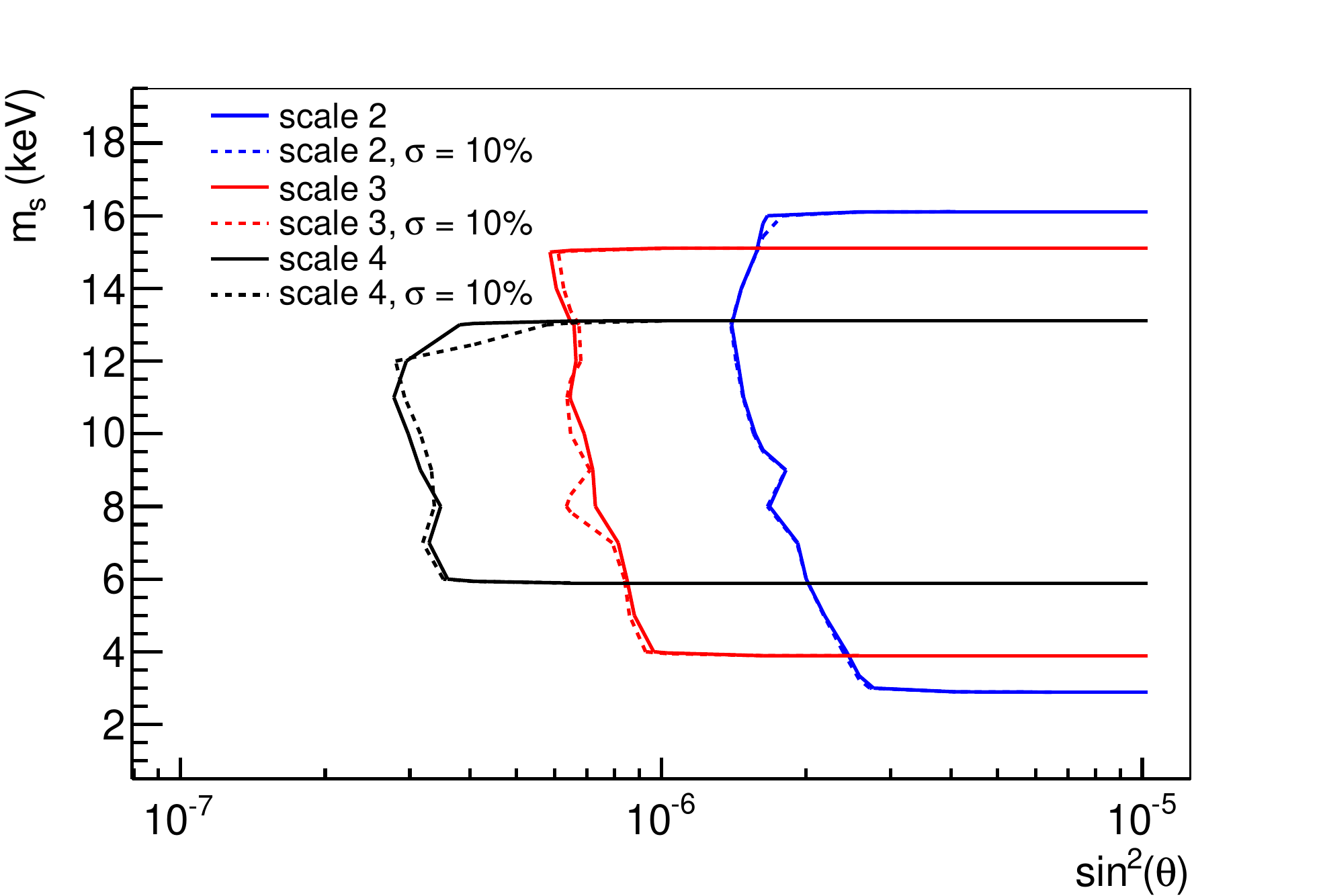} 
\caption{Exclusion limits for scale 2, 3 and 4 or a differential measurement of 3 years. The dashed lines correspond to an analysis where the spectral shape was varied with $\sigma = 10$\% as described in the main text and figure~\ref{fig:specband}. As a result we find that the exclusion limits stay largely unchanged when allowing for this 10\% uncertainty of the spectrum. This result represents the main advantage of a wavelet analysis in a keV-scale sterile neutrino search.}
\label{fig:exclusion_pol}
\end{figure}

Figure \ref{fig:exclusion_pol} shows the sensitivity contours for a spectral uncertainty of $\sigma =10\%$. It is evident that even in case of shape corrections of the order of $10\%$ to the $\beta$-decay spectrum, the sensitivity stays largely unchanged. In this range the detection of a kink-like signature with the wavelet approach is independent of the precise shape of the tritium $\beta$-decay spectrum. As long as the corrections are differentiable functions the kink can still be depicted by the wavelet. 

\subsubsection{Energy Resolution}
\begin{figure}
\includegraphics[width=0.45\textwidth]{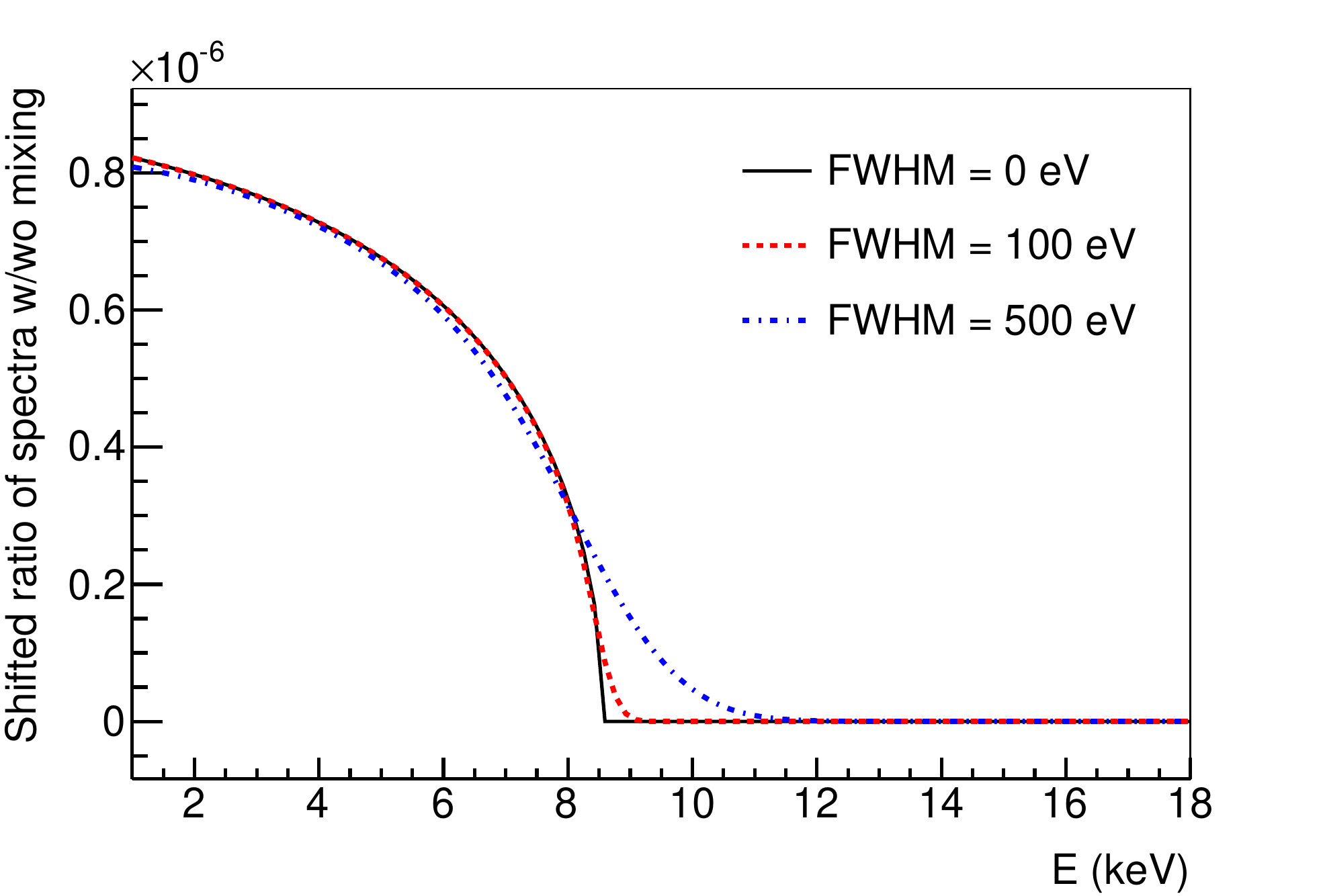}
\caption{Ratio of differential $\beta$-spectra for finite energy resolution of $\mathrm{FWHM}=0, 100, 500$~eV with $m_{\mathrm{s}}=10$ keV, $\sin^2 \theta = 10^{-6}$. The curves have been shifted to 0 beyond the kink. A finite resolution washes out the local kink signature, reducing the sensitivity of the wavelet approach.}
\label{fig:detres}
\end{figure}

\begin{figure}
\centering
\includegraphics[width=0.45\textwidth]{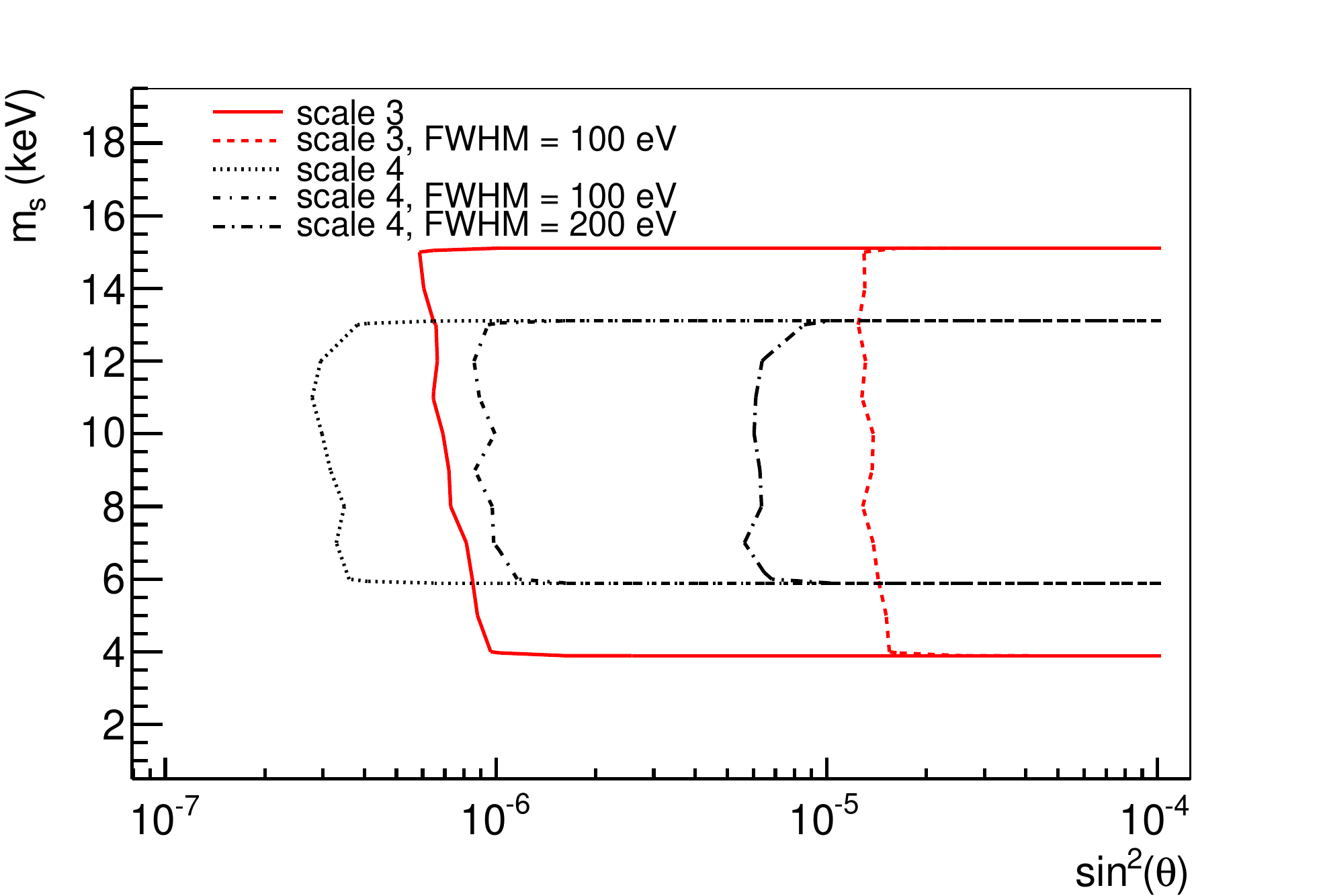} 
\caption{Influence of a finite energy resolution, leading to a washing-out to the kink-like signature for a differential measurement of 3 years. Especially for scale 3 (red solid line) the sensitivity is drastically reduced by a finite energy resolution of 100~eV (red dashed line). Scale 4 is less sensitive to a small washing-out. This is to be expected considering the intrinsic decrease of energy resolution of higher scales, as described in section~\ref{ssc:DWT}}.
\label{fig:exclusion_detres}
\end{figure}

In this section we investigate the effect of a finite energy resolution, which would wash out the local kink signature of a keV-scale sterile neutrino, as displayed in figure~\ref{fig:detres}, and thereby alter the sensitivity of the wavelet approach. To implement a washing-out of the kink signature we convolute the tritium $\beta$-decay spectrum with a Gaussian function with a certain full width half maximum (FWHM), representing the energy resolution. 

Figure \ref{fig:exclusion_detres} shows the exclusion contours for a finite energy resolution of $\mathrm{FWHM} = 100$~eV and $\mathrm{FWHM} = 200$~eV. As expected, the sensitivity is drastically reduced, since the sharp kink signature, to which the wavelet approach is sensitive, is smoothed in case of a finite resolution. We conclude that for a successful application of the wavelet approach a differential measurement with an energy resolution of the order of $\mathrm{FWHM}=100-200$~eV is necessary to achieve a sensitivity to a mixing angle of $\sin^2 \theta \sim 10^{-5}$.

In case of an integral measurement, due to the very sharp transmission characteristics of the main spectrometer, an energy resolution of 0.93~eV can be achieved~\cite{designreport,katrin}. Therefore, the energy resolution is not the critical factor for an integral measurement; the sensitivity curves in figure~\ref{fig:exclusion_int} are not modified significantly even when assuming a reduced energy resolution of the spectrometer by up to a factor of about 10.

\section{Conclusion and Outlook}
In this work we have shown that the wavelet approach is a powerful method to detect the kink-like signature of a keV-scale sterile neutrino in the tritium $\beta$-energy spectrum. With this model-independent ansatz of applying the standard DWT to a completely unmodified tritium $\beta$-decay spectrum one can reach a sensitivity of better than $\sin^2 \theta = 10^{-6}$ in a mass range of $m_{\mathrm{s}} = 4 - 15$~keV. 

Most importantly, it is demonstrated that the shape of the $\beta$-decay spectrum need not to be known to a precision level equivalent to the size of active-to-sterile neutrino mixing in order to be sensitive to the kink-signature.  As long as the corrections to the spectral shape can be described by smooth functions, the sensitivity to detect a keV-scale sterile neutrino signature remains unchanged. 

From an experimental point of view, we come to the conclusion that a KATRIN-like tritium source in combination with a differential measurement technique, offering an energy resolution of FWHM $\sim$100 eV, should allow to push forward into uncharted parameter regions of laboratory searches for sterile neutrinos. 

This sensitivity level would significantly improve on existing limits on active-to-sterile neutrino mixing in the keV-mass range from kink searches in nuclear $\beta$-decays, typically reaching levels of $\sim 10^{-3}$~\cite{Beringer}, and be the first step to probing parameter regions of interest for cosmology and astrophysics. 

\section{Acknowledgements}
S. Mertens gratefully acknowledges support by a Feodor Lynen Research fellowship of the Alexander von Humboldt Foundation, and support by the Helmholtz Association as well as by KHYS at KIT. This work was supported by the U.S. Department of Energy, Office of Science, Office of Nuclear Physics, under Contract No. DE-AC02-05CH11231. K. Dolde and M. Korzeczek would like to thank KIT for financial support and LBNL for hospitality during their internship at Berkeley. Special thanks to D.\ Radford for fruitful discussions.

\end{document}